\documentclass{article}
\usepackage{spconf,amsmath,graphicx,hyperref}
\graphicspath{{figures/}}
\usepackage{siunitx}
\usepackage{mathtools}
\usepackage{multirow, booktabs}
\usepackage{siunitx}
\usepackage{tikz}
\usepackage{pgfplots}
\usepgfplotslibrary{groupplots}
\pgfplotsset{compat=1.18}
\definecolor{okabeBlue}{RGB}{0, 114, 178}
\definecolor{okabeOrange}{RGB}{213, 94, 0}
\usepackage{cite}

\newcommand{\RR}{I\!\!R} 
\newcommand{\Nat}{I\!\!N}

\newcommand{\secref}[1]{Section~\ref{#1}}
\newcommand{\eqnref}[1]{Eq.~\eqref{#1}}
\newcommand{\figref}[1]{Fig.~\ref{#1}}
\newcommand{\tabref}[1]{Table~\ref{#1}}

\newcommand{\by}{\mathbf{y}}

\newcommand{\bX}{\mathbf{X}}

\newcommand{\bzero}{\mathbf{0}}

\title{Pitch Smoothing Using Relative Interval Networks}
\name{Chin-Yun Yu$^{\sharp}$ \qquad Chi-Jen Peng$^{\natural}$ \qquad Li Su$^{\flat}$ \qquad Gy{\"o}rgy Fazekas$^{\sharp}$}
\address{
  $^{\sharp}$ Centre for Digital Music, Queen Mary University of London, UK\\
  $^{\natural}$ Department of Information Management, National Taiwan University, Taiwan\\
  $^{\flat}$ Institute of Information Science, Academia Sinica, Taiwan
}
\begin{document}
\ninept
\maketitle
\begin{abstract}
  Pitch tracking systems typically couple a per-frame fundamental frequency ($F_0$) estimator with a temporal smoothing stage to obtain continuous trajectories.
  Conventional Viterbi smoothers enforce first-order continuity but lack long-term temporal awareness and could lock into octave errors across corrupted frames.
  We propose Relative Interval Networks (RIN), a trajectory smoothing framework that reconciles per-frame pitch estimates with data-driven multi-hop pitch differences.
  We extract robust relative pitch intervals across arbitrary frame offsets using Variable-Q Transform cross-correlation.
  We formulate pitch smoothing as an $L_1$-norm optimization problem and prove its equivalence to a minimum cost circulation problem, solved efficiently via linear programming.
  Evaluations across speech, singing, and instrumental datasets show that RIN substantially improves weak estimators, matches or outperforms Viterbi decoding at a comparable computational cost, and provides superior robustness under certain acoustic degradation.
\end{abstract}
\section{INTRODUCTION}
\label{sec:intro}

Fundamental frequency ($F_0$) estimation, commonly referred to as pitch estimation in audio signal processing, is a core task in speech and music analysis.
Numerous algorithms have been proposed, from classical methods such as YIN~\cite{de2002yin}, and SWIPE~\cite{camacho2008sawtooth}, to deep neural models like CREPE~\cite{kim2018crepe}, DeepF0~\cite{singh2021deepf0}, FCNF0++~\cite{morrisonCrossdomainNeuralPitch2023}, YOLOPitch~\cite{li2024yolopitch}, SLASH~\cite{terashima2025slash}, and PESTO~\cite{riouPESTOPitchEstimation2023}.
While modern neural models demonstrate strong resilience against noise~\cite{han2014neural,kim2018crepe}, standalone per-frame estimators cannot exploit the rich temporal structure of audio signals.
Consequently, pitch tracking, defined as the extraction of temporally coherent $F_0$ trajectories over time, typically couples front-end pitch estimation with back-end pitch smoothing~\cite{shiBayesianPitchTracking2019,zahorian2008spectral,han2014neural}.
Temporal trajectory smoothing is therefore essential to ensure physically consistent contours, which forms the focus of this work.

\begin{figure}[t]
  \centering
  \includegraphics[width=\linewidth, trim={0cm 0.1cm 0cm 0.1cm}, clip]{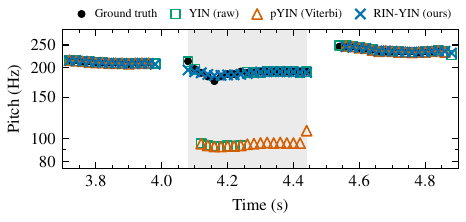}
  \vspace{-15pt}
  \caption{A failure case of Viterbi decoding on an MIR-1K excerpt.}
  \label{fig:failure_case}
\end{figure}

To enforce temporal continuity, conventional systems rely on heuristics such as median filtering~\cite{rabiner1975applications}, or on dynamic programming (DP)~\cite{huang2000dp,chen2008trues}, which minimizes a sum of local and transition costs over a discretized state space~\cite{Ney1983dp,huang2000dp,zahorian2008spectral}.
Viterbi decoding is the DP recursion for the case where these costs are the log-likelihoods of a hidden Markov model (HMM)~\cite{Viterbi1967,mauch2014pyin,han2014neural}, a combination that significantly reduces pitch errors.
Its limitations arise at three distinct layers of this construction.
The chain structure of DP penalizes only immediately adjacent frames, so consecutive corrupted frames can lock the contour into an incorrect octave, as illustrated in \figref{fig:failure_case}.
The HMM adds static, signal-agnostic transition penalties, assigning low probability to rare but legitimate transitions.
The trellis requires a fine pitch grid, which incurs substantial computational cost.

In this work, we propose Relative Interval Networks (RIN), a trajectory smoothing framework that overcomes these bottlenecks.
RIN extends our prior work on time-of-arrival estimation in spatial audio~\cite{ycy2024hrir} by adapting measurements of relative differences to pitch smoothing.
We propose a robust pitch difference estimator using the Variable-Q Transform (VQT)~\cite{schorkhuber2014matlab} and aggregating multi-hop observations beyond immediate neighbors.
By combining these pitch differences with a given per-frame pitch estimator, our framework constructs a global optimization problem that minimizes the overall discrepancy between observations from the signal.
As illustrated in \figref{fig:failure_case}, long-range relative measurements transport true pitch intervals across corrupted regions.
We also show that this problem is equivalent to a minimum cost circulation problem, for which efficient implementations exist.
Experiments on speech and musical signals show performance comparable to the Viterbi algorithm in clean conditions and superior robustness when low-SNR white noise presents, while remaining computationally efficient.
Our implementation is open source and available on GitHub~\footnote{\url{https://github.com/iamycy/rin}}.

\section{METHODOLOGY}
\label{sec:method}

\subsection{Problem formulation}
\label{ssec:problem}

Let $F_0[n] = F_0(n T_s)$ denote the true pitch contour with frame period $T_s$ in discrete time.
Throughout this work, $F_0$ is in log-scale (cents), where musical intervals and harmonic shifts are linear and additive.
At each frame $n$, a front-end pitch estimator outputs an isolated pitch estimate $\hat{F}_0[n]$ along with an associated confidence or voicing probability $v[n] \ge 0$.
Typically, a pitch trajectory evolves with a bounded maximum rate of change~\cite{xu2002maximum}.
Standard heuristic smoothers such as median filtering enforce continuity over a local window without taking frame confidences $v[n]$ into account~\cite{rabiner1975applications}.
Conversely, probabilistic methods such as HMM~\cite{mauch2014pyin} (decoded via the Viterbi algorithm) use $v[n]$ as observation likelihoods, but balance them against signal-agnostic, static transition penalties that constrain only immediate consecutive frames.

Instead of penalizing pitch differences toward zero, we propose to regularize the contour using data-driven \emph{pitch difference} estimates $\Delta_m \hat{F}_0[n] \approx F_0[n] - F_0[n-m]$ across arbitrary frame offsets $m \in \mathcal{J} \subset \Nat$.
We formulate pitch smoothing as a joint optimization problem that reconciles absolute frame estimates $\hat{F}_0[n]$ with multi-hop pitch differences $\Delta_m \hat{F}_0[n]$:
\begin{multline}
  \label{eq:optimization}
  \tilde{F}_0 = \underset{F_0}{\arg\min}  \sum_{n=0}^{N-1} v[n]\left\| F_0[n] - \hat{F}_0[n] \right\|_p \\
  + \sum_{m \in \mathcal{J}} \sum_{n=m}^{N-1} w_m[n]\left\| F_0[n] - F_0[n-m] - \Delta_m \hat{F}_0[n] \right\|_p
\end{multline}
where $w_m[n] \ge 0$ is the weights of each pitch difference estimate.

Standard smoothing approaches can be understood through \eqnref{eq:optimization}.
When the offset set is restricted to immediate neighbors ($\mathcal{J} = \{1\}$) and zero pitch differences ($\Delta_1 \hat{F}_0[n] = 0$), the second term reduces to a first-order smoothness penalty, the continuous-state analogue of the transition cost in HMM decoding under a Laplacian transition prior.
The major difference is thus the non-zero pitch differences $\Delta_m \hat{F}_0[n]$ from the acoustic signal across multi-hop offsets ($m > 1$).
\eqnref{eq:optimization} regularizes the trajectory toward measured acoustic continuity, bridging corrupted or unvoiced regions across longer temporal spans.
We set $p = 1$, as the $L_1$ norm is markedly more robust than $L_2$ against severe outliers such as octave transposition errors and resist to hard note transitions in music signals.

The proposed methodology consists of estimating pitch differences using the VQT (\secref{ssec:vqt}), deriving dynamic confidence weights (\secref{ssec:confidence_weighting}), solving the optimization problem in \eqnref{eq:optimization} using linear programming (LP) (\secref{ssec:network_flow}), as well as an extended framework to enhance voicing detection (\secref{ssec:voicing}).

\subsection{Computing pitch differences using VQT}
\label{ssec:vqt}

Let us denote a magnitude VQT frame of the target audio signal as $\bX \in \RR^K_+$, where $K$ is the total number of frequency bins, and let $X_k$ be its $k$-th element.
We choose VQT because of its shift-invariance and equal frequency resolution property: for example, if $\bX$ is pitch-shifted by $\gamma$ cents, the resulting VQT spectrum $\hat{\bX}$ has $\hat{X}_k = X_{k-\tau}$, where $\tau = \frac{B}{1200} \gamma$ and $B$ is the number of bins per octave. %
Replacing it with constant-Q transform (CQT)~\cite{Riou2025pesto} will violate this assumption as CQT have non-equal resolutions over frequencies.
For simplicity, we treat the bin shift $\tau$ as an integer here before continuous refinement.
We further assume that, across $m$ frames, the amplitudes of the harmonics remain unchanged up to a scale factor, so $X_k[n] \approx \alpha_{m}[n] X_{k-\tau}[n-m]$ where $\alpha_m[n] > 0,\, \tau = \frac{B}{1200} \Delta_m F_0[n]$.
When this assumption does not hold, we adjust $w_m[n]$ so the corresponding ill-posed estimates affect less on the overall solution, which is discussed in \secref{ssec:confidence_weighting}.

For a candidate bin shift $\tau$, a spectrum and its $\tau$-shift version overlap over $K - |\tau|$ frequency bins.
Given $\bar{\bX}_\tau[n]$ and $\bar{\bX}_{-\tau}[n-m]$ the normalized version of the slices $\bX_\tau[n]$ and $\bX_{-\tau}[n-m]$ with zero mean and unit norm, the correlation at shift $\tau$ and time index $n$ with $m$ hop is compactly expressed as the vector dot product between the two slices:
\begin{equation}
  \label{eq:cross_corr}
  r_{m,\tau}[n] = \max\left(0, \bar{\bX}_\tau[n]^\top \bar{\bX}_{-\tau}[n-m]\right),
\end{equation}
and we drop negative correlations, which indicate dissimilar matches, and keep later weights computation stay positive.
Evaluating $r_{m,\tau}[n]$ across candidate shifts $\tau \in [-\tau_{\max}, \tau_{\max}]$ yields the cross-correlation vector $\mathbf{r}_m[n] \in [0, 1]^{2\tau_{\max} + 1}$.
We first identify the maximum peak location and then apply parabolic interpolation around the peak to achieve a fine-grained continuous shift $\tilde{\tau}_m[n]$ and interpolated peak correlation $\tilde{r}_{m}[n]$ (clipped to 1 to prevent interpolation overshoot).
The final pitch difference estimate in cents is then given by $\Delta_m \hat{F}_0[n] = \frac{1200}{B} \tilde{\tau}_m[n]$.

\subsection{Confidence weighting}
\label{ssec:confidence_weighting}

The confidence weight $w_m$ reflects the reliability of the pitch difference estimates and is derived dynamically from the vector $\mathbf{r}_m$ constructed with Eq. (\ref{eq:cross_corr}).
A straightforward choice is to have $w_m=\tilde{r}_m$.
However, the peak's prominence relative to the surrounding correlations is also important: when two frames exhibit a strong harmonic match, the correlation features a prominent peak around the true shift, and significantly exceeds the average correlation across competing shifts.
Conversely, during unvoiced or noisy frames, the peak correlation is weak and barely stands out from the average.
We therefore construct the confidence weight as the product of the peak-to-mean prominence and the peak correlation:
\begin{equation}
  \label{eq:confidence_weight}
  w = \frac{2\arcsin\left(\tilde{r}\right)}{\pi} \left(1 - \frac{\bar{r}}{\tilde{r}}\right),
\end{equation}
where  $\bar{r}$ is the average correlation across all $\tau$.
For the peak correlation term, the $\arcsin$ function applies a non-linear mapping $[0, 1] \mapsto [0, 1]$ that counteracts the saturation of skewed high correlation values observed in practice.
Because $0 \le \bar{r} \le \tilde{r} \le 1$, both factors and their product $w$ are naturally bounded in $[0, 1]$.
The peak-to-mean prominence behaves similarly to spectral flatness but performed slightly better in our early experiments.
The interplay of the two terms is important.
For example, when the correlations across all $\tau$ are high, $\tilde{r}$ is also high but the peak-to-mean prominence is low and gate the false confidence, resulting in smaller weights $w$.
Evaluating \eqnref{eq:confidence_weight} on $\mathbf{r}_m[n]$ yields the relative confidence weights $w_m[n]$ for each frame $n$ and offset $m$.

\subsection{Network flow formulation}
\label{ssec:network_flow}

We adapt the formulation proposed by Yu et al.~\cite{ycy2024hrir}, starting from a graph $G = (V, E)$ where the nodes $V$ corresponds to the time frames $n \in \{0, 1, \dots, N-1\}$ plus a ground node $\delta$, and the edges $E$ consists of absolute edges $(\delta, i)$ for $i \in \{0, 1, \dots, N-1\}$ and relative edges $(n-m, n)$ for $n \in \{m, m+1, \dots, N-1\}$ and $m \in \mathcal{J}$.
Each absolute edge $(\delta, n)$ carries the observed absolute pitch $\hat{F}_0[n]$ with confidence $v[n]$, and each relative edge $(n-m, n)$ carries the observed pitch difference $\Delta_m \hat{F}_0[n]$ with confidence $w_m[n]$.
Let $F_0[\delta] = 0$ so the absolute pitch estimates are the \emph{differences} from the ground node ($\hat{F}_0[n] = \hat{F}_0[n] - F_0[\delta]$), \eqnref{eq:optimization} can be written as:
\begin{equation}
  \label{eq:lp_formulation}
  \min_{\mathbf{f}} \mathbf{w}^\top \left\lvert \mathbf{B}^\top \mathbf{f} - \boldsymbol{\Delta} \right\rvert,
\end{equation}
where $\boldsymbol{\Delta} \in \RR^{|E|}$ is the concatenation of all $\hat{F}_0$ and $\Delta_m \hat{F}_0$, $\mathbf{w} \in \RR^{|E|}$ is the concatenation of corresponding edge weights $\{v, w_m\}$, and $\mathbf{f} \in \RR^N$ is the vector of $F_0$ we wish to estimate.
$\mathbf{B} \in \RR^{N \times |E|}$ is the incidence matrix of the graph $G$ omitting $\delta$, where each column corresponds to an edge and has a $-1$ at the row of the source node and a $1$ at the row of the target node, and zeros elsewhere.

Instead of solving \eqnref{eq:lp_formulation} directly, there is a dual formulation which can be solved more efficiently.
Specifically, the weighted $L_1$ norm can be expressed as a maximization over its dual variable using Fenchel duality~\cite{boyd2004convex}:
\begin{equation}
  \mathbf{w}^\top \left\lvert \mathbf{B}^\top \mathbf{f} - \boldsymbol{\Delta} \right\rvert = \max_{-\mathbf{w} \le \by \le \mathbf{w}} \by^\top \left(\mathbf{B}^\top \mathbf{f} - \boldsymbol{\Delta} \right),
\end{equation}
where $\by \in \RR^{|E|}$ denotes the circulation flow along the edges of $G$.
Interchanging the minimization over unconstrained $\mathbf{f}$ and the maximization over $\by$ requires $\mathbf{B}\by = \bzero$ to prevent the inner $\min_{\mathbf{f}} \mathbf{f}^\top(\mathbf{B}\by)$ from diverging to $-\infty$.
This condition enforces Kirchhoff's flow conservation law at every node.
Negating the objective yields the equivalent minimum cost circulation problem:
\begin{equation}
  \label{eq:dual_circulation}
  \begin{gathered}
    \min_{\by} ~ \boldsymbol{\Delta}^\top \by \\
    \text{s.t.} \quad \mathbf{B}\by = \bzero, \quad -\mathbf{w} \le \by \le \mathbf{w}.
  \end{gathered}
\end{equation}
By LP strong duality~\cite{boyd2004convex}, the optimal pitch contour $\tilde{\mathbf{f}}$ is recovered directly as the Lagrange multipliers (node potentials) associated with the flow conservation constraints $\mathbf{B}\by = \bzero$~\cite{ahuja1993network}.
In practice, these are the dual values of the equality constraints returned by the LP solver, and no second solve is needed.
Solving \eqnref{eq:dual_circulation} directly shrinks the linear system from $|E|$ constraints over $N + 2|E|$ variables in standard primal slack form~\cite{ma2022timeseriesphase} down to $N$ constraints over $|E|$ bounded variables.
In our experiments, it roughly halves the solve time using HiGHS~\cite{huangfu2018parallelizing}.
Applying network flow theory to solve the global pitch trajectory motivated us to name the proposed method Relative Interval Networks (RIN).

\subsection{Enhancing voicing detection}
\label{ssec:voicing}
In addition to pitch estimation, it is crucial to determine whether each frame contains voiced or unvoiced audio content.
Instead of replacing the original voicing detection $v[n]$ entirely, we propose to enhance it by incorporating additional cues derived from our pitch difference estimation framework.
The confidence weight $w_m[n]$ derived from \secref{ssec:confidence_weighting} is an indirect indicator of this: a frame with higher correlation weights $w_m[n]$ indicates stronger and more consistent pitch patterns across frames, so it is likely to be voiced.
We thus propose a measure for voicing detection:
\begin{equation}
  s[n] = \sqrt{\frac{\sum_{m \in \mathcal{J}_n^+} \beta_m w_m^2[n] + \sum_{m \in \mathcal{J}_n^-} \beta_m w_m^2[n+m]}{\sum_{m \in \mathcal{J}_n^+} \beta_m + \sum_{m \in \mathcal{J}_n^-} \beta_m}},
\end{equation}
where $\mathcal{J}_n^+ = \{m \in \mathcal{J} : n \ge m\}$ and $\mathcal{J}_n^- = \{m \in \mathcal{J} : n+m < N\}$ denote the sets of valid backward and forward offsets, respectively, and $\beta_m > 0$ are the mixing weights, for which we found $\beta_m = \frac{1}{m^2}$ works the best.
The enhanced voicing strength is calculated as the geometric mean of the two: $\tilde{v}[n] = \sqrt{v[n] s[n]}$. 
Taking 0.5 as the fixed reference threshold for $s[n]$ and $\lambda$ as the front-end threshold, the new voicing detection threshold is $\sqrt{\lambda/2}$.

\section{Experiments and results}
\label{sec:experiments}

\begin{table*}[t]
  \centering
  \setlength{\tabcolsep}{2.2pt}
  \caption{Clean audio pitch tracking on PTDB-TUG, MIR-1K, and URMP, with RPA, RCA, and OA in \% and MAE in cents. Bold marks the best result among four configurations in each non-runtime column. Runtime is measured on MIR-1K in ms/s.}
  \label{tab:clean_benchmark}
  \small
  \renewcommand{\arraystretch}{0.7}
  \begin{tabular}{ll cccc @{\hspace{4pt}\vrule\hspace{4pt}} cccc @{\hspace{4pt}\vrule\hspace{4pt}} cccc @{\hspace{8pt}} r}
    \toprule
                       &                             & \multicolumn{4}{c}{\textbf{PTDB-TUG}} & \multicolumn{4}{c}{\textbf{MIR-1K}} & \multicolumn{4}{c}{\textbf{URMP}} & \textbf{Runtime}                                                                                                                                                                                                                                                      \\
    \cmidrule(lr){3-6} \cmidrule(lr){7-10} \cmidrule(lr){11-14} \cmidrule(lr){15-15}
    \textbf{Estimator} & \textbf{Smoothing}          & \textbf{RPA} $\uparrow$               & \textbf{RCA} $\uparrow$             & \textbf{MAE} $\downarrow$         & \textbf{OA} $\uparrow$ & \textbf{RPA} $\uparrow$ & \textbf{RCA} $\uparrow$ & \textbf{MAE} $\downarrow$ & \textbf{OA} $\uparrow$ & \textbf{RPA} $\uparrow$ & \textbf{RCA} $\uparrow$ & \textbf{MAE} $\downarrow$ & \textbf{OA} $\uparrow$ & \textbf{ms/s} $\downarrow$ \\
    \midrule
    \multirow{4}{*}{YIN}
                       & --                          & 62.36                                 & 67.72                               & \textbf{11.34}                    & 89.19                  & 94.90                   & 95.95                   & \textbf{7.01}             & 90.25                  & 87.56                   & 91.14                   & \textbf{2.33}             & 90.45                  & 2.1                        \\
                       & Viterbi (pYIN)                    & 83.31                                 & 85.48                               & 13.09                             & 89.76                  & \textbf{98.64}          & \textbf{99.06}          & 7.40                      & 90.72                  & \textbf{95.15}          & \textbf{97.85}          & 3.89                      & 90.58                  & 13.3                       \\
                       & RIN $\mathcal{J}_A$   & 82.96                                 & 84.87                               & 12.91                             & 90.23                  & 97.11                   & 97.81                   & 7.67                      & 91.15                  & 93.93                   & 97.35                   & 2.98                      & 90.93                  & 5.2                        \\
                       & RIN $\mathcal{J}_B$ & \textbf{84.20}                        & \textbf{86.07}                      & 13.09                             & \textbf{90.24}         & 97.42                   & 97.76                   & 8.46                      & \textbf{91.33}         & 94.43                   & 97.64                   & 3.38                      & \textbf{91.22}         & 5.5                        \\
    \midrule
    \multirow{4}{*}{SWIPE}
                       & --                          & \textbf{85.46}                        & 86.92                               & \textbf{12.90}                    & \textbf{92.79}         & 96.39                   & 96.92                   & 9.37                      & 90.46                  & 94.34                   & 95.38                   & 5.88                      & 92.15                  & 49.2                       \\
                       & Viterbi                     & 84.88                                 & 86.92                               & 13.23                             & 92.68                  & 96.36                   & 96.54                   & 9.79                      & 90.32                  & 95.70                   & 96.53                   & 6.84                      & 92.63                  & 50.7                       \\
                       & RIN $\mathcal{J}_A$   & 85.20                                 & 86.94                               & 13.09                             & 92.62                  & 97.26                   & 97.55                   & 9.22                      & 90.69                  & 95.68                   & 96.65                   & 5.69                      & 92.80                  & 52.3                       \\
                       & RIN $\mathcal{J}_B$ & 84.79                                 & \textbf{87.20}                      & 13.26                             & 92.52                  & \textbf{97.53}          & \textbf{97.70}          & \textbf{9.09}             & \textbf{90.87}         & \textbf{95.97}          & \textbf{96.70}          & \textbf{5.46}             & \textbf{92.90}         & 52.7                       \\
    \midrule
    \multirow{4}{*}{CREPE}
                       & --                          & 85.91                                 & 88.10                               & \textbf{12.96}                    & 92.19                  & 97.57                   & 97.91                   & 10.36                     & 89.47                  & 94.41                   & 96.25                   & 7.00                      & 92.22                  & 1851.0                     \\
                       & Viterbi                     & \textbf{86.18}                        & \textbf{89.25}                      & 13.21                             & 92.04                  & \textbf{97.81}          & 97.84                   & 10.45                     & 89.58                  & 95.55                   & \textbf{97.46}          & 7.27                      & 92.62                  & 1852.8                     \\
                       & RIN $\mathcal{J}_A$   & 85.81                                 & 87.69                               & 13.01                             & \textbf{92.99}         & 97.71                   & \textbf{97.96}          & 9.88                      & 90.56                  & 95.06                   & 96.78                   & 6.85                      & 92.78                  & 1854.1                     \\
                       & RIN $\mathcal{J}_B$ & 85.72                                 & 87.67                               & 13.02                             & 92.90                  & 97.68                   & 97.83                   & \textbf{9.40}             & \textbf{90.62}         & \textbf{95.73}          & 96.94                   & \textbf{6.24}             & \textbf{93.07}         & 1854.4                     \\
    \midrule
    \multirow{4}{*}{FCNF0++}
                       & --                          & \textbf{93.31}                        & \textbf{94.05}                      & \textbf{10.57}                    & \textbf{93.39}         & 91.27                   & 93.06                   & 11.91                     & 78.09                  & 62.84                   & 67.06                   & 8.72                      & 51.60                  & 189.8                      \\
                       & Viterbi                     & 91.56                                 & 93.73                               & 11.03                             & 93.32                  & 92.21                   & 92.92                   & 12.19                     & 78.08                  & 72.55                   & 76.48                   & 9.80                      & 51.71                  & 191.7                      \\
                       & RIN $\mathcal{J}_A$   & 91.84                                 & 92.57                               & 11.67                             & 92.61                  & 93.41                   & 94.53                   & 11.39                     & 86.51                  & 77.23                   & 80.92                   & 8.14                      & 60.19                  & 192.8                      \\
                       & RIN $\mathcal{J}_B$ & 88.80                                 & 90.10                               & 12.87                             & 92.26                  & \textbf{95.24}          & \textbf{95.73}          & \textbf{10.71}            & \textbf{87.03}         & \textbf{83.60}          & \textbf{86.34}          & \textbf{7.66}             & \textbf{60.27}         & 193.2                      \\
    \midrule
    \multirow{4}{*}{PESTO}
                       & --                          & 82.96                                 & 86.27                               & \textbf{14.25}                    & 89.16                  & \textbf{98.26}          & \textbf{98.50}          & 11.96                     & 96.10                  & 95.00                   & 96.54                   & 9.36                      & 40.04                  & 35.8                       \\
                       & Viterbi                     & \textbf{84.58}                        & \textbf{87.96}                      & 14.63                             & 89.37                  & 98.25                   & 98.32                   & 11.98                     & 96.10                  & 94.53                   & 96.38                   & 9.99                      & 40.05                  & 37.1                       \\
                       & RIN $\mathcal{J}_A$   & 80.91                                 & 84.04                               & 14.26                             & \textbf{89.80}         & 98.10                   & 98.30                   & 10.73                     & \textbf{96.46}         & \textbf{95.67}          & \textbf{97.02}          & 8.68                      & \textbf{47.34}         & 38.8                       \\
                       & RIN $\mathcal{J}_B$ & 81.64                                 & 84.62                               & 14.30                             & \textbf{89.80}         & 97.95                   & 98.07                   & \textbf{9.59}             & 96.38                  & 95.45                   & 96.53                   & \textbf{7.96}             & 47.33                  & 39.2                       \\
    \bottomrule
  \end{tabular}
\end{table*}

\subsection{Experimental setup and metrics}
\label{subsec:exp_setup}

We evaluate on three pitch-annotated datasets.
\textbf{MIR-1K}~\cite{hsu2010mir1k} contains 1,000 isolated vocal tracks from Chinese pop songs.
\textbf{PTDB-TUG}~\cite{pirker2011ptdb} contains 4,718 clips of read speech.
\textbf{URMP}~\cite{li2019urmp} contains multi-instrument classical ensemble pieces, from which we use the 149 separately recorded monophonic \texttt{AuSep} tracks.
We consider the following pitch estimators as baselines: YIN~\cite{de2002yin}, SWIPE~\cite{camacho2008sawtooth}, CREPE~\cite{kim2018crepe}, FCNF0++~\cite{morrisonCrossdomainNeuralPitch2023}, and PESTO~\cite{riouPESTOPitchEstimation2023}.
Each is evaluated in four configurations: without smoothing, Viterbi, and RIN with $\mathcal{J}_A=\{1\}$ and $\mathcal{J}_B=\{1,5\}$.
We use $\mathcal{J}_A$ to match Viterbi's adjacent-frame neighborhood, while $\mathcal{J}_B$ adds one longer offset to test the benefit of the multi-hop structure, for which we empirically found 5 to work well.

All configurations use a \SI{20}{\ms} frame period.
All Viterbi baselines use the pYIN decoder of \texttt{librosa}~\cite{mauch2014pyin,librosa10} at its default settings, except for a maximum transition rate of 25 octaves per second, matching our $\tau_{\max}$.
Under Viterbi, YIN follows the pYIN implementation in \texttt{librosa}, with pitch and voicing decoded jointly.
For the other estimators, candidate scores are interpolated to the same grid as \texttt{librosa}'s pYIN, with an additional softmax transform for SWIPE, before pitch-only decoding as in \texttt{torchcrepe}; native voicing strengths are retained.
YIN uses a 0.15 selection threshold for candidates and voicing strength $v[n]=1-e[n]$, with $e[n]$ the selected candidate's normalized difference~\cite{de2002yin}.
RIN uses a 252-bin VQT at 36 bins per octave, selected in a preliminary sweep in which both coarser and finer resolutions reduced accuracy, while keeping other parameters to \texttt{librosa}'s default.
We empirically select $\tau_{\max}$ to be 18 after visualizing a few samples from MIR-1K, which limits $\Delta_m F_0[n]$ to be within $\pm$6 semitones and is strictly above human limits when $m\le 2$~\cite{xu2002maximum}.

We compute Raw Pitch Accuracy (RPA), Raw Chroma Accuracy (RCA), and Overall Accuracy (OA) at a 50-cent tolerance using \texttt{mir\_eval}~\cite{raffel2014mireval}, define MAE as the mean absolute cent error over frames with the same tolerance, and report runtime.
RPA, RCA, and MAE use unthresholded pitch contours, which means evaluating on all the voiced reference pitch, whereas OA is the fraction of frames that are correctly unvoiced or voiced within 50 cents.
We compute all four metrics per clip and macro-average them.
We set the voicing detection threshold $\lambda$ to 0.5 by default, except for SWIPE, where we found 0.3 works better. 
The thresholds for RIN is elaborated in \secref{ssec:voicing}.
Runtime is sequential computation only, measured on Ubuntu using a single CPU thread pinned to one core of an Intel Core i9-14900K processor, over all MIR-1K clips in milliseconds and divided by the total audio duration.

\vspace{-5pt}
\subsection{Benchmark performance on clean audio}
\label{subsec:clean_results}
\tabref{tab:clean_benchmark} compares RIN at $\mathcal{J}_A$ and $\mathcal{J}_B$ with Viterbi and unsmoothed estimates.
RIN's clearest advantage is its ability to recover RPA and RCA after domain-specific front-end failures, as seen for FCNF0++ on URMP and YIN on PTDB-TUG.
Across MIR-1K and URMP, RIN generally matches or outperforms Viterbi in RPA and tends to yield lower MAE when their RPA is comparable. 
On PTDB-TUG, the unsmoothed SWIPE and FCNF0++ estimates retain the highest RPA, and neither smoother reduces MAE relative to the unsmoothed estimate. 

Averaged across estimators within each dataset, RIN at $\mathcal{J}_B$ generally yields a smaller gap between RPA and RCA than Viterbi on PTDB-TUG (2.1 versus 2.6 points) and URMP (1.8 versus 2.2), indicating fewer octave confusions; the methods have similar gaps on MIR-1K, both below $0.3$. 
RIN gives higher OA than Viterbi for every estimator on MIR-1K and URMP, while remaining within about one percentage point on PTDB-TUG.
Across most estimator-dataset pairs, $\mathcal{J}_A$ and $\mathcal{J}_B$ yield similar RPA and RCA and remain broadly competitive with Viterbi. 
Despite these similar accuracies, adding offset 5 generally lowers MAE on MIR-1K and URMP. 
RIN has comparable runtime to Viterbi for most estimators. 
For $\mathcal{J}_B$, the LP stage takes 0.77-0.83\,ms/s.
Because the VQT and front-end estimator are independent, running them in parallel would leave only the slower component followed by the LP stage on the critical path, further reducing runtime.

\subsection{Robustness and ablation studies}

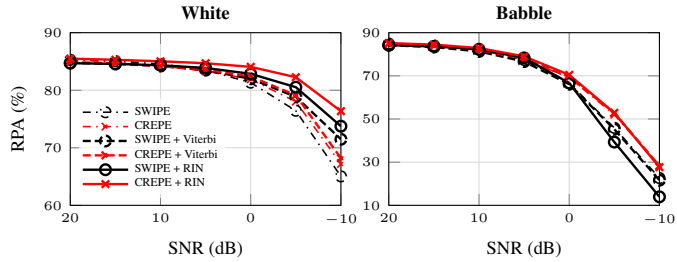
\begin{figure}[t]
  \centering
  \begin{tikzpicture}
  \begin{groupplot}[
      group style={
          group size=2 by 1,
          horizontal sep=18pt,
        },
      width=0.60\linewidth,
      height=0.45\linewidth,
      x dir=reverse,
      xmin=-10, xmax=20,
      xtick={20, 10, 0, -10},
      xticklabels={$20$, $10$, $0$, $-10$},
      xlabel={SNR (dB)},
      grid=both,
      grid style={line width=0.4pt, draw=gray!30},
      tick label style={font=\tiny},
      label style={font=\scriptsize},
    ]
    \nextgroupplot[
      title={\textbf{White}},
      title style={font=\scriptsize, yshift=-4pt},
      ylabel={RPA (\%)},
      ymin=60, ymax=90,
      ytick={60, 70, 80, 90},
      legend style={
          at={(0.03, 0.04)},
          anchor=south west,
          font=\fontsize{4.5pt}{5.5pt}\selectfont,
          draw=none,
          fill=white,
          fill opacity=0.7,
          text opacity=1,
          cells={anchor=west},
          inner sep=1pt,
          row sep=-3pt,
          legend columns=1,
        },
    ]
    \addplot[
      color=black,
      line width=0.6pt,
      dashdotted,
      mark=o,
    ] table [x=snr, y=swipe_raw] {figures/noise_white.dat};
    \addlegendentry{SWIPE}

    \addplot[
      color=red,
      line width=0.6pt,
      dashdotted,
      mark=x,
    ] table [x=snr, y=crepe_raw] {figures/noise_white.dat};
    \addlegendentry{CREPE}

    \addplot[
      color=black,
      line width=0.9pt,
      dashed,
      dash pattern=on 3pt off 1.5pt,
      mark=o,
      mark size=2pt,
    ] table [x=snr, y=swipe_vit] {figures/noise_white.dat};
    \addlegendentry{SWIPE + Viterbi}

    \addplot[
      color=red,
      line width=0.9pt,
      dashed,
      dash pattern=on 3pt off 1.5pt,
      mark=x,
      mark size=2pt,
    ] table [x=snr, y=crepe_vit] {figures/noise_white.dat};
    \addlegendentry{CREPE + Viterbi}

    \addplot[
      color=black,
      line width=0.9pt,
      mark=o,
      mark size=2pt,
    ] table [x=snr, y=swipe_rin] {figures/noise_white.dat};
    \addlegendentry{SWIPE + RIN}

    \addplot[
      color=red,
      line width=0.9pt,
      mark=x,
      mark size=2pt,
    ] table [x=snr, y=crepe_rin] {figures/noise_white.dat};
    \addlegendentry{CREPE + RIN}

    \nextgroupplot[
      title={\textbf{Babble}},
      title style={font=\scriptsize, yshift=-4pt},
      ymin=10, ymax=90,
      ytick={10, 30, 50, 70, 90},
    ]
    \addplot[
      color=black,
      line width=0.6pt,
      dashdotted,
      mark=o,
    ] table [x=snr, y=swipe_raw] {figures/noise_babble.dat};

    \addplot[
      color=red,
      line width=0.6pt,
      dashdotted,
      mark=x,
    ] table [x=snr, y=crepe_raw] {figures/noise_babble.dat};

    \addplot[
      color=black,
      line width=0.9pt,
      dashed,
      dash pattern=on 3pt off 1.5pt,
      mark=o,
      mark size=2pt,
    ] table [x=snr, y=swipe_vit] {figures/noise_babble.dat};

    \addplot[
      color=red,
      line width=0.9pt,
      dashed,
      dash pattern=on 3pt off 1.5pt,
      mark=x,
      mark size=2pt,
    ] table [x=snr, y=crepe_vit] {figures/noise_babble.dat};

    \addplot[
      color=black,
      line width=0.9pt,
      mark=o,
      mark size=2pt,
    ] table [x=snr, y=swipe_rin] {figures/noise_babble.dat};

    \addplot[
      color=red,
      line width=0.9pt,
      mark=x,
      mark size=2pt,
    ] table [x=snr, y=crepe_rin] {figures/noise_babble.dat};

  \end{groupplot}
\end{tikzpicture}
  \caption{RPA under additive white and babble noise at various SNRs on PTDB-TUG.}
  \vspace{-15pt}
  \label{fig:noise_robustness}
\end{figure}

\figref{fig:noise_robustness} compares the unsmoothed, Viterbi, and RIN at $\mathcal{J}_B$ on PTDB-TUG mixtures with white and babble noise from \SIrange[]{20}{-10}{\decibel} SNR.
Under white noise, RIN's gain over Viterbi increases as the SNR decreases.
At \SI{-10}{\decibel}, RIN with CREPE outperforms Viterbi at every SNR and reaches an RPA of 76.36, compared with 68.19 for Viterbi and 67.18 without smoothing; for SWIPE, RIN reaches 73.75, compared with 71.48 for Viterbi.
Under babble noise, RIN performs closely to Viterbi for CREPE across all SNRs, and matches or outperforms Viterbi at \SI{0}{\decibel} and above for SWIPE, but falls below both Viterbi and the unsmoothed estimate at negative SNR.
This contrast is consistent with white noise being temporally uncorrelated, whereas competing harmonics in babble can mislead pitch difference estimates.

\begin{figure}[t]
  \centering
  \begin{tikzpicture}
  \begin{axis}[
      width=0.7\linewidth,
      height=0.45\linewidth,
      xmin=0, xmax=7,
      ymin=64, ymax=79,
      xtick={0, 1, 2, 3, 4, 5, 6, 7},
      ytick={65, 70, 75},
      yticklabel style={/pgf/number format/fixed, /pgf/number format/precision=0},
      xlabel={$\lvert \mathcal{J} \rvert$},
      ylabel={RPA (\%)},
      grid=both,
      grid style={line width=0.4pt, draw=gray!30},
      tick label style={font=\scriptsize},
      label style={font=\footnotesize},
      unbounded coords=discard,
      legend style={
          at={(0.97, 0.05)},
          anchor=south east,
          font=\scriptsize,
          draw=none,
          fill=none,
          cells={anchor=west, align=left},
          inner sep=0pt,
          row sep=3pt,
        },
    ]
    \addplot[
      color=black, line width=0.9pt,
      mark=triangle, mark size=2pt,
      mark indices={1, 2, 3, 4, 6, 7, 8},
      restrict x to domain=0:7,
    ] table [x=k, y=full_oracle] {figures/hop_ablation.dat};
    \addlegendentry{SWIPE greedy oracle}

    \addplot[
      color=red, line width=0.9pt,
      mark=square, mark size=1.8pt,
      mark indices={1, 2, 3, 4, 5, 7, 8},
      restrict x to domain=0:7,
    ] table [x=k, y=full_oracle] {figures/hop_ablation_crepe.dat};
    \addlegendentry{CREPE greedy oracle}

    \addplot[only marks, mark=star, mark size=3.5pt, line width=1.1pt, color=black, forget plot]
    coordinates {(4, 74.8459)};

    \addplot[only marks, mark=star, mark size=3.5pt, line width=1.1pt, color=red, forget plot]
    coordinates {(5, 77.8926)};

    \addplot[only marks, mark=diamond*, mark size=2.8pt, color=black, forget plot]
    coordinates {(2, 73.75)};

    \addplot[only marks, mark=diamond*, mark size=2.8pt, color=red, forget plot]
    coordinates {(2, 76.36)};
  \end{axis}
\end{tikzpicture}
    \caption{Ablation of differential offset sets on PTDB-TUG under additive white noise at \SI{-10}{\decibel} SNR.}
  \label{fig:hop_ablation}
\end{figure}
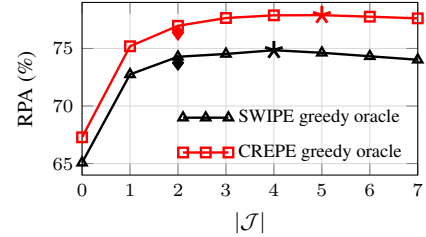

\figref{fig:hop_ablation} traces greedy oracle paths on PTDB-TUG under white noise at \SI{-10}{\decibel} SNR, where each step adds the offset from 1 to 15 with the largest RPA gain.
The paths therefore serve only as in-sample references, while the diamonds mark $\mathcal{J}_B$. %
The first offset provides most of the improvement, raising RPA from 65.07 to 72.75 for SWIPE and from 67.28 to 75.19 for CREPE; one-offset SWIPE already surpasses unsmoothed CREPE, and gains largely plateau beyond two offsets.
At $\lvert\mathcal{J}\rvert = 2$, $\mathcal{J}_B$ is only 0.53 and 0.61 points below the greedy sets $\{2,7\}$ and $\{2,9\}$ for SWIPE and CREPE, respectively.
The full paths peak at 74.85 with $\{2,3,7,15\}$ for SWIPE and 77.89 with $\{2,3,5,9,15\}$ for CREPE, 1.10 and 1.53 points above $\mathcal{J}_B$.
The distinct peak sets suggest that the choice of offsets depends on the estimator.

\vspace{-5pt}
\section{CONCLUSION}
We introduced RIN, a network flow pitch smoother that combines per-frame pitch estimates with multi-hop pitch difference estimates using VQT and their confidence weights.
Across speech, singing, and instrumental benchmarks, RIN improved YIN by up to 21.8 points in RPA, but did not consistently benefit stronger estimators.
RIN's advantage over Viterbi increased as the SNR decreased under white noise, while under babble noise it remained comparable except for SWIPE at the lowest SNRs.
Offset ablations showed that effective offsets depend on the estimator.
Concurrent VQT feature computing would leave the measured LP stage as the only subsequent stage, which is below 1 ms/s 
when the offsets are $\{1, 5\}$.
Future work includes exploring pitch differences in other $F_0$ domains besides log-frequency, and investigating more efficient solvers for the network flow optimization.

\section{ACKNOWLEDGMENT}
Yu is a research student at the UKRI CDT in AI and Music, jointly supported by UKRI (grant EP/S022694/1) and QMUL.
Fazekas was supported by the Leverhulme Trust and the Royal Academy of Engineering under the RAEng / Leverhulme Trust Research Fellowships scheme.
Claude Opus 5 and Copilot with Gemini 3.7/3.8 were used to conduct the experiments and help edit the initial draft.

\section{COMPLIANCE WITH ETHICAL STANDARDS}
This research study was conducted retrospectively using human subject data made available in open access by MIR-1K, PTDB-TUG, and URMP.
Ethical approval was thus not required.

\bibliographystyle{IEEEbib}
\bibliography{refs}

\end{document}